\documentclass[prl,twocolumn,superscriptaddress,showpacs,amsmath,amssymb]{revtex4}
\usepackage{amsmath}
\usepackage{graphicx}
\usepackage{dcolumn}

\begin{document}
\title{Confinement effects on intermediate state flux patterns in mesoscopic type-I superconductors}
\author{G. R. Berdiyorov}
\affiliation{Department of Physics, University of Antwerp,
Groenenborgerlaan 171, B-2020 Antwerpen, Belgium}
\author{A. D. Hernandez}
\affiliation{Centro Atomico Bariloche, 8400 San Carlos de Bariloche,
Rio Negro, Argentina }
\author{F.~M.~Peeters}
\email{francois.peeters@ua.ac.be} \affiliation{Department of
Physics, University of Antwerp, Groenenborgerlaan 171, B-2020
Antwerpen, Belgium}

\date{ \today }

\date{\today}

\begin{abstract}

Intermediate state (IS) flux structures in mesoscopic type-I
superconductors are investigated within the Ginzburg-Landau theory.
In addition to well-established tubular and laminar structures, the
strong confinement leads to the formation of (i) a phase of
\textit{singly quantized vortices}, which is typical for type-II
superconductors and (ii) \textit{a ring} of a normal domain at
equilibrium. The stability region and the formation process of these
IS flux structures are strongly influenced by the geometry of the
sample.

%In contrast to bulk type-I superconductors, where flux tubes
%represent the equilibrium topology of the IS [R. Prozorov, Phys.
%Rev. Lett. {\bf 98}, 257001 (2007)], both tubular and laminar
%structures form the ground state of mesoscopic samples.

%For example, laminae are predominantly located parallel to the
%surface of a superconducting cube while they are radially oriented
%in a sphere.

%The size of bubbles is found to be independent of the long-range
%interaction between the normal state domains. We calculate the
%equilibrium diameter of an isolated bubble resulting from the
%competition between the Biot-and-Savart interaction of the Meissner
%current encircling the bubble and the superconductor-normal
%interface energy. The isolation of each bubble in the superconductor
%and the interface energy are shown to preclude any continuous size
%variation of the bubbles after their formation, contrary to the
%prediction of mean-field models.

\end{abstract}


\maketitle

Intermediate state (IS) of type-I superconductors has received a
revival of interest in recent years \cite{Fidler,pro2} as one of the
rare systems where competing interactions lead to the formation of
spatially modulated structures. This kind of complex structures have
been observed in other systems like ferrofluids, amphiphilic
monolayers and chemical reaction-diffusion systems \cite{Seul}.
Analogies to type-I superconductors also extend to astrophysics
\cite{Buckley}.

IS flux structures can be described theoretically assuming parallel
stripes of normal and superconducting (SC) domains \cite{landau} or
a periodic array of multiquanta flux tubes \cite{goren}. The tubular
structure is more mobile, while the laminar structure is
topologically constrained \cite{Hoberg} and transition between these
states occurs for larger volume fraction of normal domains
\cite{cebers}. However, these regular patterns are rarely seen in
experiment \cite{huebener}. Instead, much more complex flux
structures -- highly branched and intricate fingered patterns of
flux domains -- are often observed
\cite{huebener,faber,miller,kiendl,pro1,pro2}. In addition, type-I
superconductors show clear hysteresis -- flux tubes are obtained
during magnetic field penetration and laminar structures are formed
during magnetic field expulsion \cite{pro1,Velez}, suggesting that
the system gets stuck in a metastable state, i.e., the sample is not
in thermodynamic equilibrium. Zero-bulk pinning \cite{pro2}
superconducting spheres and cones show no hysteresis with flux tubes
dominating the IS. With increasing applied magnetic field, a new
phase -- a suprafroth -- is formed \cite{Fidler}, which represents
neither tubular nor laminar structure.

%\textit{Therefore, which of those two structures -- flux tubes or
%laminar patterns -- actually form the ground state of the IS is
%still an open question.}
%These observations let to conclude \cite{pro2,Hoberg} that flux
%tubes represent the equilibrium state of the IS.
%However, calculations within the current-loop model \cite{cebers}
%predict a transition from the flux tube state to the laminar state
%for larger volume fraction of the N domains.

The behavior becomes richer in the mesoscopic regime, where, in
addition to the competition between the magnetic energy that favors
the formation of small normal domains and the positive surface
energy that tends to form large domains, confinement effects become
important. In type-II superconductors confinement strongly
influences the distribution of vortices leading to, e.g., the
formation of giant vortices \cite{kanda} and symmetry induced
vortex-antivortex pairs \cite{chibo}. Misko et al., stabilized such
vortex-antivortex patterns in a long type-I superconducting prism
with triangular cross section \cite{slava}, due to the competition
between vortex-antivortex repulsion and finite size effects.
However, their study was limited to a very narrow range of applied
magnetic field. Our main goal in this paper is a systematic study of
ground state flux structures in mesoscopic type-I superconductors.
We also address another important problem not yet studied neither
experimentally nor theoretically: the effect of sample geometry on
these mesoscopic IS patterns.

%We found that in addition to tubular and laminar patterns
%\textit{single quantized vortices} and \textit{a ring of a N domain}
%can be stabilized at equilibrium, which are never found in bulk
%samples.

%Up to now, the ground state flux structures in mesoscopic type-I
%superconductors have not been investigated.
%, and use the link variable approach \cite{kato}

From the theoretical point of view, the difficulty of modeling the
IS comes from the 3D nature of the magnetic interaction of the
domains. Therefore, approximate expressions are usually used for the
magnetic energy. \cite{goren,goldstein,dorsey,cebers} In our
approach we used the Ginzburg-Landau (GL) theory, where \textit{no
approximation} for the magnetic energy is used and the structure of
the domains is \textit{not predetermined}. We solve the GL equations
for the order parameter $\Psi$ and the vector potential ${\bf A}$:
\begin{eqnarray}
&& (-i\nabla-{\bf A})^2\Psi = (1-|\Psi|^2)\Psi, \label{gl1}\\
&& -\kappa^2\nabla\times\nabla\times {\bf A} =
\textrm{Im}(\Psi^*\nabla\Psi)-|\Psi|^2{\bf A}, \label{gl2}
\end{eqnarray}
where $\kappa$ is the GL parameter. Here, the distance is measured
in units of the coherence length $\xi$, the vector potential
$\vec{A}$ in $c\hbar/2e\xi$, and the order parameter $\Psi$ in
$\sqrt{-\alpha/\beta}$ with $\alpha$, $\beta$ being the GL
coefficients. Following the numerical approach of Ref. \cite{schw},
we apply a finite-difference representation for the order parameter
and the vector potential on a uniform 3D Cartesian space grid (up to
$256\times256\times256$ grid points). $\Psi$ satisfies the boundary
condition $(-i\nabla-{\bf A})\Psi |_n=0$ at the sample surface and
far away from the superconductor ${\bf A}$ is determined by the
external applied field. The simulations of the IS are conducted in
i) field sweep up: we started from the full Meissner state
($|\Psi|=1$) and slowly increased the magnetic field, after reaching
the stationary state; ii) field sweep down: we started simulations
with $|\Psi|=0$ and $H>H_c$ and decreased the field with small
steps; and iii) field cooled simulations starting from random
initial conditions for each value of the applied field.

\begin{figure}[t] \centering
\vspace{0cm}
\includegraphics[width=\linewidth]{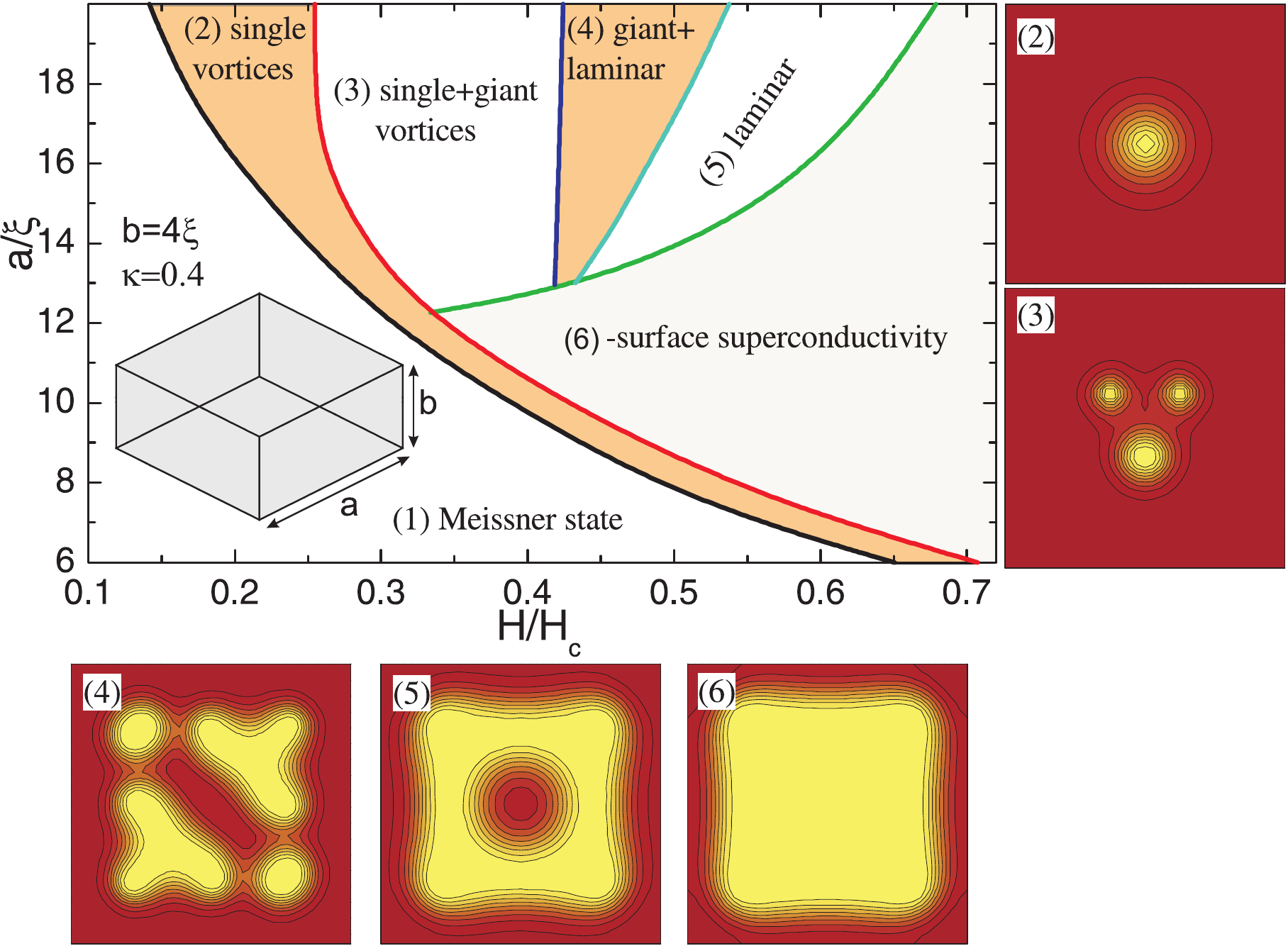}
\vspace{0cm} \caption{\label{figure1}(Color online) The ground state
flux structures for a rectangular prism of size $a$ and height $b$
(see the inset) as a function of $a$ and the applied magnetic field
$H$ for $b=4\xi$ and $\kappa=0.4$. The insets (2-6) denote the
Cooper-pair density plots [red (dark gray)/yellow (light gray)
corresponds to high/low $|\psi|^2$] of the corresponding states.}
\end{figure}

\textbf{\textit{Equilibrium flux structures}}. -- As a
representative example of mesoscopic type-I superconductors, we
studied IS flux structures in a rectangular prism of size $a$ and
height $b$ exposed to a homogeneous field $H$ (see insets of Fig.
\ref{figure1}). We constructed the equilibrium phase diagram as a
function of $a$ and $H$ for $\kappa=0.4$ and $b=4\xi$ as shown in
Fig. \ref{figure1}. The ground state was found from the field cooled
simulations by comparing the energies of all found states for given
magnetic field and sample parameters. The Meissner state (region 1)
is found at small fields and the stability of this phase increases
to larger fields with decreasing sample size. The most prominent
feature of this phase diagram is the existence of region 2 where
\textit{singly quantized fluxes} (i.e., individual vortices) are
nucleated in the sample (see inset 2). Note that this state is
typical for type-II superconductors and was never found in bulk
type-I samples \cite{hernandez,pro1,pro2}, where the energy of
singly quantized vortices is larger than the energy of multiply
quantized flux tubes. At higher fields individual vortices are too
close to each other and giant vortices nucleate together with
individual vortices (region 3) as illustrated by inset 3 of Fig.
\ref{figure1}. In region 4 laminar-like structures start to form in
combination with flux tubes (inset 4). Such laminar-like structures
are obtained for larger $a$, usually in the form of {\it a ring} of
normal domain (inset 5). This novel state is a consequence of strong
confinement imposed by the sample surface and resembles recently
found suprafroth in macroscopic SC spheres \cite{Fidler}. Surface
superconductivity (region 6) is formed before the system transits to
the normal state (inset 6). Notice also that, we did not find
branching of normal domains, which occurs in very thick samples
\cite{huebener}. From this phase diagram we notice that the ground
state flux structures in mesoscopic type-I superconductors do not
depend only on the applied magnetic field (as in the case of bulk
superconductors \cite{cebers}), but also on the confinement due to
the finite size of the sample. The latter leads to the formation of
singly quantized flux tubes and a ring of normal domain.

\begin{figure}[b] \centering
\includegraphics[width=\linewidth]{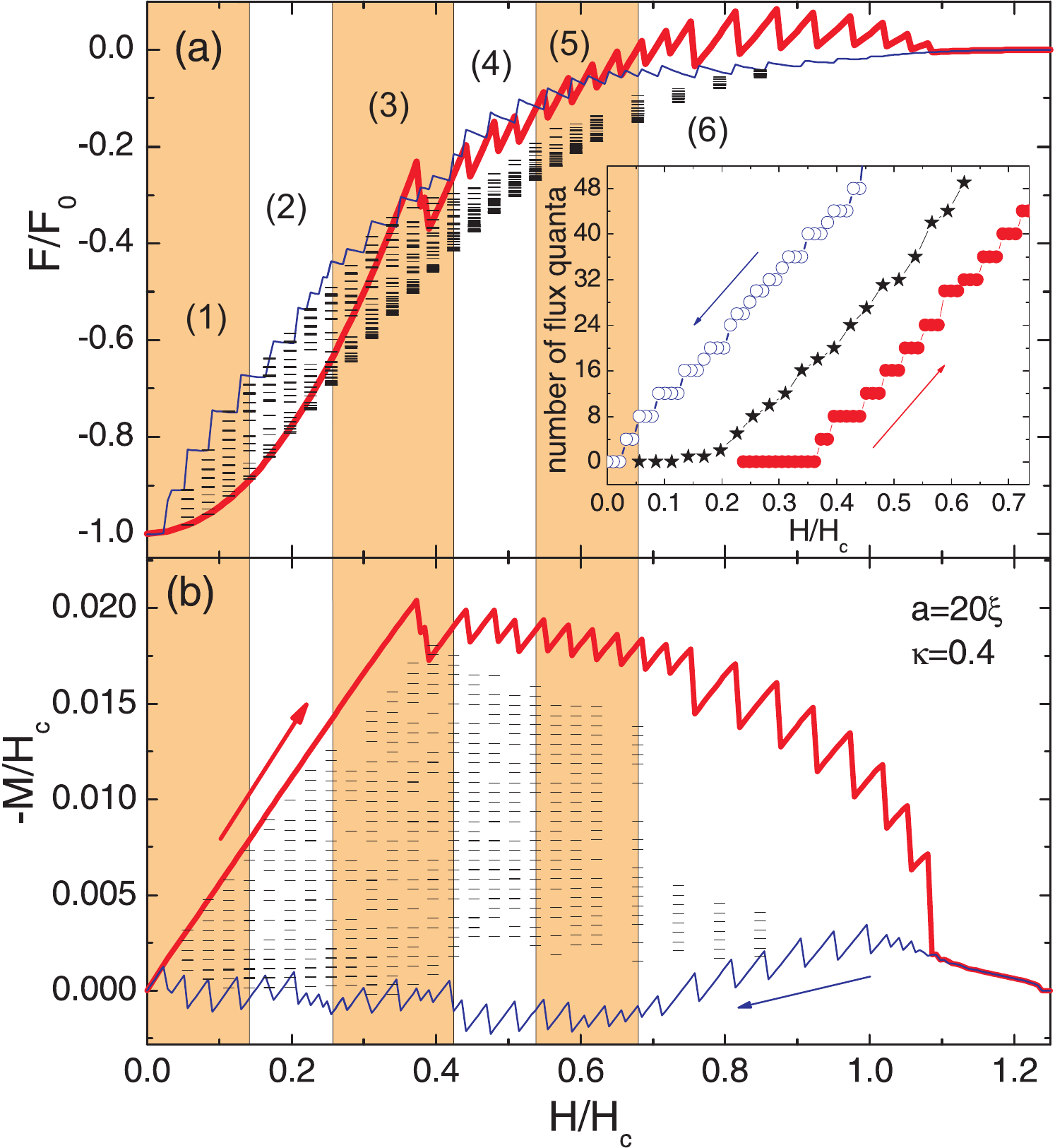}
\vspace{0cm} \caption{\label{figure2}(Color online) The free energy
(a) and the magnetization (b) as a function of the applied magnetic
field for a superconducting cube of size $a=20\xi$ and for
$\kappa=0.4$. Thick red (blue thin) curves are the results obtained
when increasing (decreasing) the magnetic field and the symbols
(hyphens) show the results of field cooling simulations. The inset
shows the total number of flux in the sample as a function of the
applied field obtained during field sweep up (filled red circles),
field sweep down (open blue circles) and of the lowest energy state
from field cooled (stars) simulations. }
\end{figure}
%See the text for a description of the different stability regions of
%the ground state (regions 1 to 6).

\begin{figure}[b] \centering
\includegraphics[width=\linewidth]{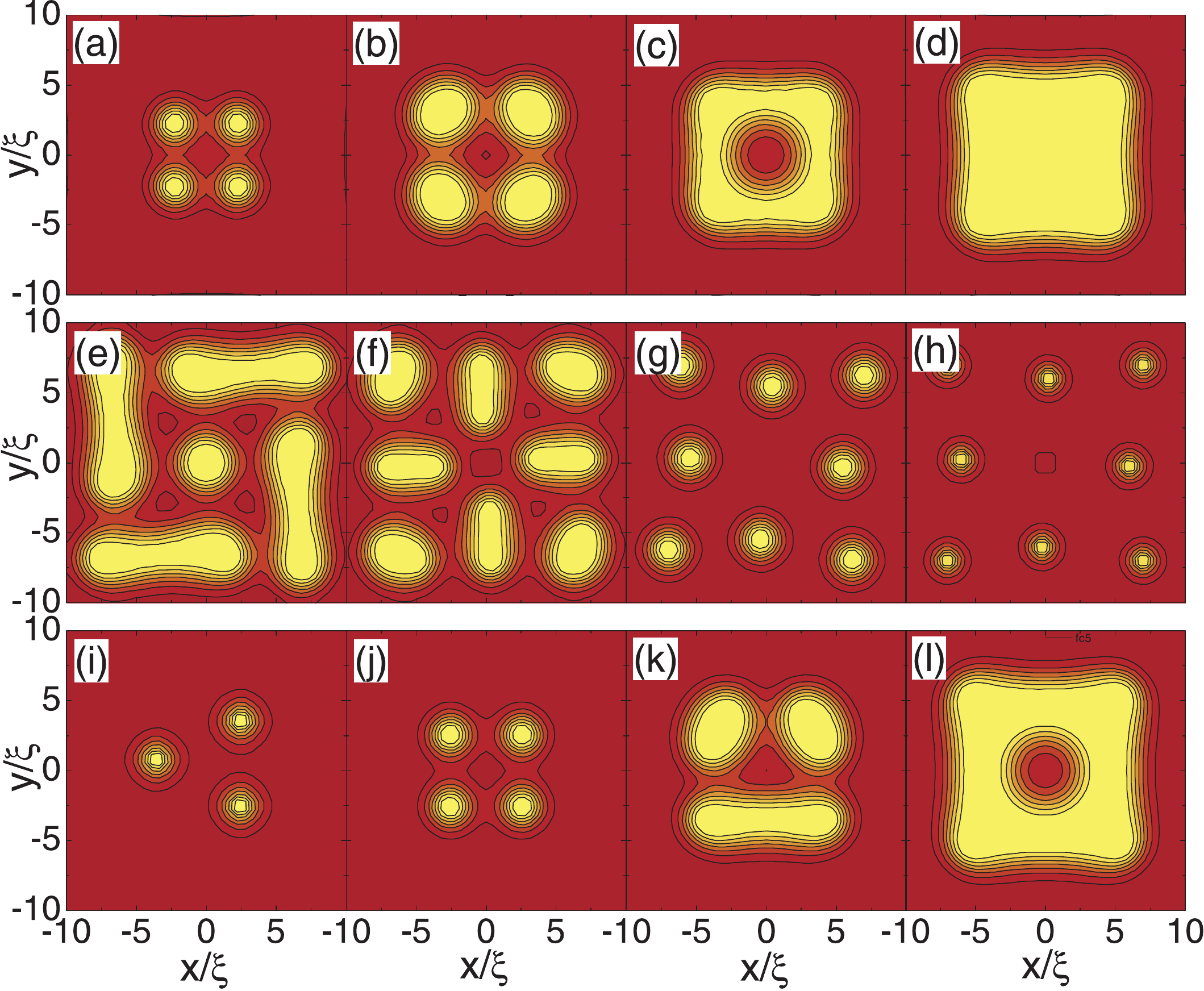}
\vspace{0cm} \caption{\label{figure3}(Color online) Contour plots of
the Cooper pair density $|\psi|^2$ in the middle plane of the cube
obtained during magnetic field sweep up (a-d), down (e-h) and in
field cooling regime (the ground state) (i-l) at the applied fields
$H/H_c$=: 0.44 (a), 0.52 (b), 0.55 (c), 0.66 (d), 0.43 (e), 0.37
(f), 0.15 (g), 0.08 (h), 0.17 (i), 0.28 (j), 0.48 (k), and 0.57 (l).
The number of fluxoids $N$ is: 8 (a), 20 (b), 28 (c), 36 (d), 48
(e), 40 (f), 16 (g), 8 (h), 3 (i), 8 (j), 22 (k), and 40 (l). }
\end{figure}

%Another important problem not addressed neither experimentally nor
%theoretically in type-I superconductors is the study of the IS in
%samples of sizes comparable with the expected periodicity of the
%patterns, while interesting "mesoscopic" behaviors have been found
%in type-II superconductors with sizes of the order of few times the
%magnetic size of vortices.21

\textbf{\textit{Effect of sample geometry}}. An important factor
that influences the formation of IS patterns is the sample shape and
geometry \cite{pro2,fortini,castro}. For example, a recent
experiment showed \cite{pro2} that zero-bulk pinning discs and slabs
show hysteretic behavior -- flux tubes appear on magnetic field
penetration and lamellae on flux exit. Spheres and cones show no
hysteresis with flux tubes dominating the intermediate field region.
In what follows, we study the influence of the sample geometry on
the IS flux structures, as an example of a superconducting cube and
sphere with the same superconducting volume $V$. Figure
\ref{figure2} shows the free energy (in units of $F_0=H_c^2V/8\pi$)
(a) and the magnetization (b) of a cube with size $a=20\xi$ as a
function of the applied magnetic field. In the field sweep up regime
(thick red curve), the sample is in the Meissner state up to
$H_p=0.38H_{c}$ -- the penetration of flux is prohibited by the
surface barrier. When the penetration field $H_p$ is reached four
flux quanta enter the sample forming one circular flux tube or a
giant vortex. With a small increase of the field, 4 flux quanta
enter forming 4 giant vortices, each with vorticity 2 (Fig.
\ref{figure3}(a)). The strong confinement in the mesoscopic regime
prevents the formation of hexagonal structures and we usually obtain
square symmetric structures (Figs. \ref{figure3}(a,b,e-h)). Previous
studies showed that with increasing field magnetic flux enters the
sample in the form of tubes \cite{pro1} which were believed to be
due to the presence of a surface barrier \cite{fortini}. The
smallness of our sample enhances the effect and only a single flux
quantum enters through each side of the cube and join the existing
normal domains. In this way, the square symmetry of the flux
structure remains unaltered and only the number of flux quanta in
each normal domain increases (Fig. \ref{figure3}(b)). With further
increasing the field a ring of normal phase is formed surrounding a
SC domain (Fig. \ref{figure3}(c)). This ring structure closes at
$H=0.66H_{c}$ and only surface superconductivity (Fig.
\ref{figure3}(d)) remains until the system transits to the normal
state.

%Note also the existence of superheated states with energies larger
%than the normal state energy (see Fig. \ref{figure2}).

The topological hysteresis in the IS flux structures in macroscopic
samples \cite{pro2} was previously explained by the presence of a
surface barrier which is absent in decreasing fields \cite{brandt}
-- a large number of lamellae are connected to the sample edge
during flux exit, allowing the magnetic flux to exit continuously,
while these large normal domains are broken into smaller pieces
during flux penetration. Thin blue curve in Fig. \ref{figure2}(a)
shows the free energy of the cube when decreasing the magnetic field
$H$. Surface superconductivity first nucleates resulting in a narrow
flux-free zone, as obtained in the experiment \cite{valko}. With
further decreasing $H$, a SC phase nucleates in the central part of
the sample, embedded within the normal phase (similar to the state
in Fig. \ref{figure3}(c)) reducing the total number of flux quanta
(see the inset of Fig. \ref{figure2}(a)). At $H=0.43H_{c}$ the long
normal domain breaks down into 4 smaller domains that are arranged
into a square symmetric structure (Fig. \ref{figure3}(e)). When
reducing the magnetic field further the length of the normal
lamellae decreases until they are reduced to flux tubes (Fig.
\ref{figure3}(f)). The expulsion of flux occurs in such a way that
only part of the tubes leave the sample and the number of normal
domains remain unchanged (Fig. \ref{figure3}(g)). The expulsion of
the flux occurs with smaller steps compared to the case of magnetic
field sweep up. Further decreasing the field leads to the formation
of patterns containing {\it singly-quantized vortices} (Fig.
\ref{figure3}(h)). At $H=0.023H_{c}$ the magnetic field is totally
expelled and the system transits to the Meissner state. Notice that
the square symmetry of the IS flux structures is always preserved,
while in large samples the square symmetry is broken when decreasing
the field \cite{hernandez}. This difference is a consequence of
stronger confinement (surface barrier) in mesoscopic samples.

%\begin{figure}\centering
%\vspace{0cm}
%\includegraphics[width=\linewidth]{fig5}
%\vspace{0cm} \caption{\label{figure5}(Color online) Isosurface plots
%of $|\psi|^2$ of a superconducting sphere of radius $R=12.4\xi$ at
%the magnetic fields: $H=0.23H_{c}$ (N=3) (a), $H=0.42H_{c}$ (N=29)
%(b), $H=0.53H_{c}$ (N=30) (c), $H=0.57H_{c}$ (N=32) (d),
%$H=0.62H_{c}$ (N=34) (e) and $H=0.7H_{c}$ (N=36) (f). }
%\end{figure}

From Fig. \ref{figure2}(a) it is clear that the ground state of the
sample is not reached during both field sweep up and down (except
for the Meissner phase) -- the system is locked in higher energy
metastable states. The ground state can be obtained during field
cooling simulations, the results of which are shown by symbols in
Fig. \ref{figure2}. As seen from this figure the system can nucleate
in many different (meta)stable states when starting from different
random initial conditions. The total number of flux quanta $N$
inside the sample is smaller during magnetic field sweep up (solid
circles) and it is much larger when decreasing the field (open
circles) as compared to the one in the ground state (stars) (see the
inset of Fig. \ref{figure2}(a)). The existence of such superheated
and supercooled states results in a large hysteresis in the
magnetization curve \cite{gourdon}. The magnetization of the sample,
defined as the flux expelled from the sample $M=(\langle
H\rangle-H)/4\pi$, where $\langle H\rangle$ is averaged over the
sample volume magnetic field, \cite{schw} is shown in Fig.
\ref{figure2}(b). Each jump in the magnetization curve corresponds
to a transition between different flux states. The difficulty of
expelling flux results in a positive magnetization in decreasing
field (thin blue curve in Fig. \ref{figure2}(b)). The hysteresis
disappears at very small fields, and at larger magnetic fields when
only surface superconductivity exists. Transition from the tubular
state to the laminar-like state (4) occurs at $H$$\simeq$$0.42H_c$,
which is slightly larger than the field ($H\simeq0.3H_c$)
corresponding to the transition between tubular and laminar states
in bulk samples \cite{cebers}. The corresponding ground state flux
structures are shown in Figs. \ref{figure3}(i-l).

\begin{figure}[t]\centering
\vspace{0cm}
\includegraphics[width=\linewidth]{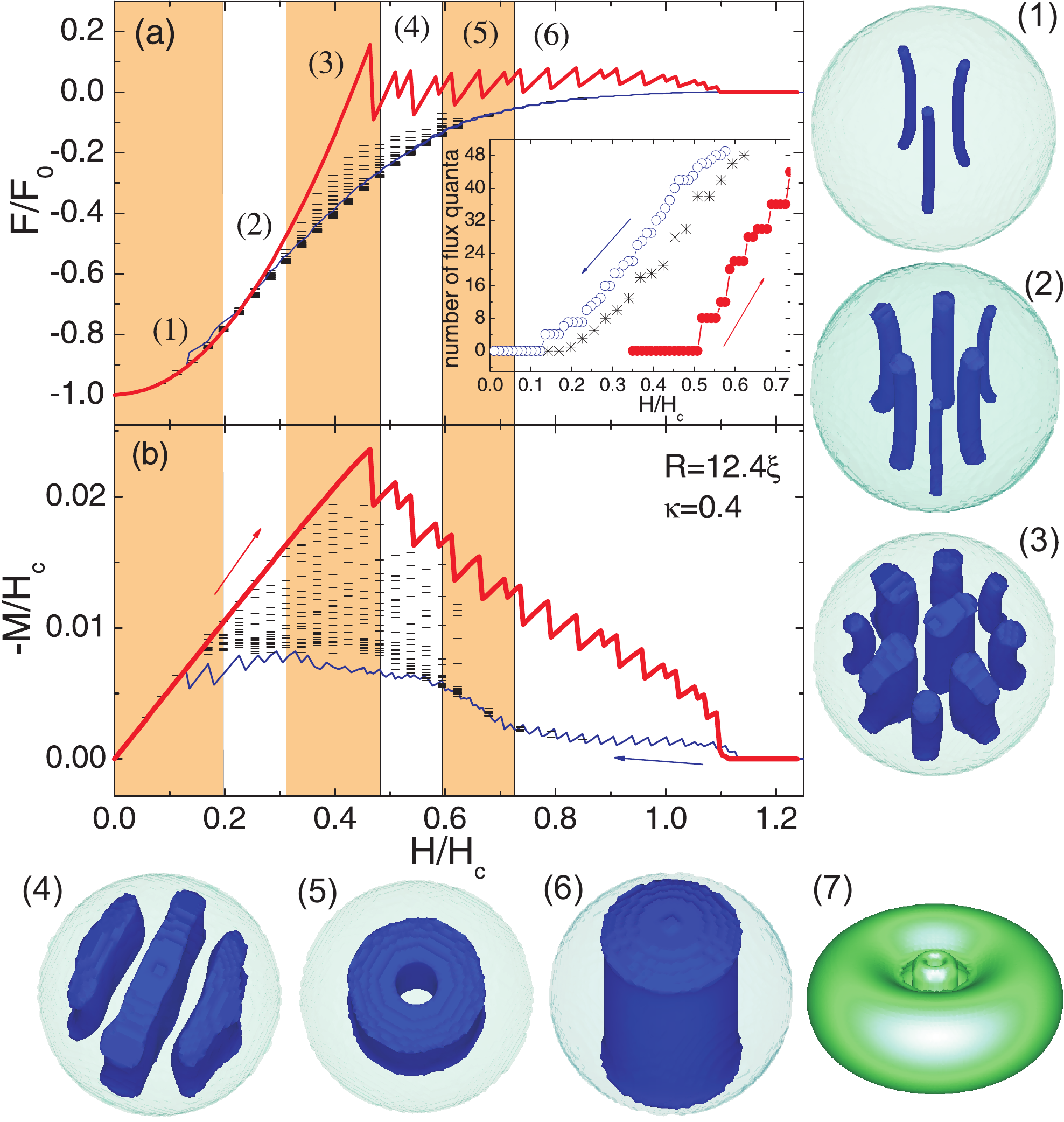}
\vspace{0cm} \caption{\label{figure4}(Color online) The same as Fig.
\ref{figure2}, but for a sphere of radius $R=12.4\xi$. Isosurface
plots show the ground state flux structures in the sphere at the
magnetic fields: $H=0.23H_{c}$ (N=3) (1), $H=0.31H_{c}$ (N=10) (2),
$H=0.51H_{c}$ (N=38) (3), $H=0.59H_{c}$ (N=46) (4), $H=0.62H_{c}$
(N=48) (5) and $H=0.73H_{c}$ (N=55) (6). Light (dark) blue
corresponds to superconducting (normal) region. Inset 7 shows the
magnetic field distribution around the sphere corresponding to the
state shown in inset 5.}
\end{figure}

Next, we contrast the results of the cube to the case of a
mesoscopic sphere. Figure \ref{figure4} shows the free energy (a)
and magnetization (b) vs. the magnetic field $H$ for a sphere of
radius $R=12.4\xi$. The effect of the sample geometry leads to the
following differences. First, the Meissner effect becomes more
pronounced leading to a larger first penetration field
($H_p=0.47H_{c}$) (thick red curve) \cite{fortini,castro}.
%This isalso a consequence of the smaller demagnetization factor of the
%sphere ($\eta_s=0.53$) compared to the one of the cube
%($\eta_c=0.62$), which makes the IS region smaller (i.e., larger
%$H_p$). Note that $\eta_s$ is larger than the one for a macroscopic
%sphere ($\eta_s=1/3$) \cite{huebener}.
Due to the nonzero demagnetization factor of the sphere the magnetic
field is strongly disturbed near the sample boundary, resembling the
magnetic field profile of a dipole (see inset 7 in Fig.
\ref{figure4}). The jump in the total number of flux quanta becomes
larger as compared to the case of the cube (compare filled circles
in the insets of Figs. \ref{figure2}(a) and \ref{figure4}(a)).
Second, the number of possible meta(stable) states (hyphens in Fig.
\ref{figure4}(a)) is substantially smaller in field cooling
simulations, especially at low and high fields. Only at midrange
fields a larger number of metastable states is found, but energy
levels are very close to each other. The most interesting results
are found for the magnetic field sweep down regime: the jumps in the
free energy (and magnetization) curve (thin blue curve) are much
smaller, leading to an almost continuous change in the number of
flux quanta. Moreover, the system follows very closely the ground
state, with only a few more flux quanta trapped inside the sample
(compare open circles and stars in the inset of Fig.
\ref{figure4}(a)). As a result of all this, the magnetic hysteresis
is smaller and no paramagnetic effect is observed (Fig.
\ref{figure4}(b)). These results allow us to conclude that the
surface barrier for penetration of magnetic field in a sphere is
larger compared to the one in flat samples (i.e., a cube), while it
is strongly reduced during flux expulsion. However, in contrast to
the macroscopic sphere \cite{Velez}, there is no complete
disappearance of the surface barrier in our mesoscopic sphere.
Isosurface plots (1-6) in Fig. \ref{figure4} shows the ground state
flux structures of the sphere for different magnetic field values.
At small fields only singly quantized vortices are found in the
ground state (inset 1) and, with increasing $H$, giant vortices
appear (inset 2). Due to the boundary condition the flux tubes
should be perpendicular to the surface of the sample, leading to the
curvature of flux tubes. Laminar-like structures first appear in the
form of deformed flux tubes (inset 3) and radially oriented stripes
(inset 4) or a ring of normal domain (inset 5). And, finally, we
arrive at the surface superconductivity state with one big central
normal domain (inset 6).

Concluding, a remarkable variety of possible flux structures in the
IS of type-I superconductors are found in the mesoscopic regime. For
example, single-quantized flux tubes (vortices) are stabilized due
to the confinement imposed by the sample surface. The latter also
leads to the formation of a ring of normal domain at larger applied
fields. All these findings are summarized into the phase diagram
(Fig. \ref{figure1}). The transition point between different flux
structures depends on the geometry of the sample which is related to
the surface barrier for flux entry and exit. Regardless of the
sample geometry the ground state of the IS at low fields consists of
flux tubes and the transition to the laminar-like state takes place
at larger magnetic fields.

% as compared to the bulk case.

%\section{Acknowledgments}

This work was supported by the Belgian Science Policy (IAP) and the
collaborative project FWO-MINCyT (FW/08/01). G.R.B. acknowledges
support from FWO-Vlaanderen.

\end{document}